\documentclass[12pt,a4paper]{article}
\usepackage{fullpage}
\usepackage{amssymb,amsmath,stmaryrd}
\usepackage[mathscr]{eucal}
\usepackage{epsfig}
\usepackage{pstricks,pst-node,pst-text,pst-3d}

\newtheorem{definition}{Definition}
\newtheorem{theorem}{Theorem}
\newtheorem{proposition}{Proposition}

\newtheorem{lemma}{Lemma}

\newcommand{\Q}{\mathcal{Q}}

\newcounter{remark}

\newcounter{example}

\newenvironment{proof}{\medskip\noindent \bf Proof: \rm}{\hspace*{\fill}
$\blacksquare$ \newline \medskip}  

\begin{document}

\title{Derivative of functions over lattices as a basis for the notion of
interaction between attributes}

\author{Michel GRABISCH\\
Universit\'e Paris I - Panth\'eon-Sorbonne\\
\normalsize email \texttt{Michel.Grabisch@lip6.fr}
\and
Christophe LABREUCHE\\
Thales Research \& Technology\\
Domaine de Corbeville, 91404 Orsay Cedex, France\\
\normalsize email \texttt{Christophe.Labreuche@thalesgroup.com}}

\date{}

\maketitle

\begin{abstract}
The paper proposes a general notion of interaction between attributes, which can
be applied to many fields in decision making and data analysis. It generalizes
the notion of interaction defined for criteria modelled by capacities, by
considering functions defined on lattices. For a given problem, the lattice
contains for each attribute the partially ordered set of remarkable points or
levels. The interaction is based on the notion of derivative of a function
defined on a lattice, and appears as a generalization of the Shapley value or
other probabilistic values.
\end{abstract}

\section{The concept of interaction: an introduction}
\label{sec:intro}
Let us consider a set $N$ of criteria describing the preferences of a decision
maker (DM) over a set $X$ of objects, alternatives, etc. We assume that for any
object $x\in X$, we are able to build a vector of \emph{scores}
$(a_1,\ldots,a_n)$ describing the satisfaction of the DM for $x$, w.r.t. each
criterion. For this reason, and in order to remain at an abstract level, we call
this vector a \emph{tuple}, which we identify with the object or alternative. We
may suppose for the moment that scores are given on the real interval $[0,1]$,
with 0 and 1 having the meaning of ``unacceptable'' and ``totally satisfying''
respectively.

We make the simplifying assumption that the preference of the DM is solely
determined by \emph{binary} tuples, i.e. whose scores are either 0 or 1 on each
criterion, the preference for other tuples being more or less an interpolation
between binary tuples. More precisely, denoting by $(1_A,0_{A^c})$ the binary
tuple having a score of 1 for all criteria in $A\subseteq N$, and 0 elsewhere,
this amounts to assigning an overall score $v(A)$ in $[0,1]$ to $(1_A,0_{A^c})$.
Doing this for all $A\subseteq N$, we have defined a set function
$v:2^N\longrightarrow[0,1]$.

Although this is not essential in the sequel, we may impose to $v$ some natural
properties. First, we may set $v(\emptyset):=0$ and $v(N):=1$, since
$A=\emptyset$ (resp. $N$) corresponds to a binary act having all its scores
being equal to 0 (resp. 1). Second, considering $A\subseteq B$, this leads to
two binary tuples of which one \emph{dominates} the other, in the sense that on
each criterion one is at least as good as the other. Then it seems natural to
impose $v(A)\leq v(B)$. This is called \emph{isotonicity}. A set function $v$
satisfying these two conditions is called a \emph{capacity} \cite{cho53} (also
called \emph{fuzzy measure} \cite{sug74}).

Let us now consider the case $n=2$ in some detail. There are 4 binary tuples
$(0,0)$, $(0,1)$, $(1,0)$ and $(1,1)$, and we know already that the first and
last have overall scores $0=v(\emptyset)$ and $1=v(N)$. What about the 2
remaining ones~? There are two extreme situations, under isotonicity.
\begin{itemize}
\item $v(\{1\})=v(\{2\})=0$, which means that for the DM, both criteria have to
be satisfactory in order to get a satisfactory tuple, the satisfaction of only
one criterion being useless. We say that the criteria are \emph{complementary}.
\item $v(\{1\})=v(\{2\})=1$, which means that for the DM, the satisfaction of
one of the two criteria is sufficient to have a satisfactory tuple, satisfying
both being useless. We say that the criteria are \emph{substitutive}. 
\end{itemize}
Clearly, in these two situations, the criteria are not independent, in the
sense that the satisfaction of one of them acts on the usefulness of the other
in order to get a satisfactory tuple (necessary in the first case, useless in
the second). So we may say that there is some \emph{interaction} between the
criteria\footnote{For further discussion on substitutive and complementary
  criteria, see Marichal \cite{mar00b}.}. 

What should be a situation where no interaction occurs, i.e. criteria act
independently~? It is a situation where the satisfaction of each criterion
brings its own contribution to the overall satisfaction, hence:
\[
v(\{1,2\}) = v(\{1\}) + v(\{2\}).
\]
Note that in the first situation, $v(\{1,2\})>v(\{1\}) + v(\{2\})$, while the
reverse inequality holds in the second situation. This suggests that the
interaction $I_{12}$ between criteria 1 and 2 should be defined as~:
\begin{equation}
\label{eq:int}
I_{12}:= v(\{1,2\}) -v(\{1\}) - v(\{2\})+v(\emptyset).
\end{equation}
This is simply the difference between binary tuples on the diagonal (where
there is strict dominance) and on the anti-diagonal (where there is no dominance
relation). The interaction is positive when criteria are complementary, while it
is negative when they are substitutive. This is consistent with intuition
considering that when criteria are complementary, they have no value by
themselves, but put together they become important for the DM.

In the case of more than 2 criteria, the definition of interaction is more
tricky but follows the same idea (see below). In fact, when $n>2$, we may
define the interaction between $3, 4,\ldots,n$ criteria as well. The general
definition of interaction for capacities has been given in \cite{gra96f}, and
has been axiomatized in \cite{grro97a}. 

\medskip

\begin{figure}[htb]
\begin{center}
\setlength{\unitlength}{0.6cm}
\begin{picture}(10,10)
        \put(1,1){\line(1,0){8}}
        \put(1,9){\line(1,0){8}}
        \put(1,1){\line(0,1){8}}
        \put(9,1){\line(0,1){8}}
        \put(0.5,5){\vector(1,0){9}}
        \put(5,0.5){\vector(0,1){9}}
        \put(5,9){\circle*{0.25}}
        \put(9,9){\circle*{0.25}}
        \put(5,5){\circle*{0.25}}
        \put(9,5){\circle*{0.25}}
        \put(1,9){\circle{0.25}}
        \put(1,5){\circle{0.25}}
        \put(1,1){\circle{0.25}}
        \put(5,1){\circle{0.25}}
        \put(9,1){\circle{0.25}}
        \put(5.1,8.5){\small $(0,1)$}
        \put(7.7,8.5){\small $(1,1)$}
        \put(5.1,5.2){\small $(0,0)$}
        \put(7.7,5.2){\small $(1,0)$}
        \put(5.1,1.2){\small $(0,-1)$}
        \put(7.3,1.2){\small $(1,-1)$}
        \put(1.1,8.5){\small $(-1,1)$}
        \put(1.1,5.2){\small $(-1,0)$}
        \put(1.1,1.2){\small $(-1,-1)$}
        \put(8.1,9.3){\small $v(\{1,2\},\emptyset)$}
        \put(4.1,9.6){\small $v(\{2\},\emptyset)$}
        \put(0,9.3){\small $v(\{2\},\{1\})$}
        \put(8.1,0.3){\small $v(\{1\},\{2\})$}
        \put(4.1,0.1){\small $v(\emptyset,\{2\})$}
        \put(0,0.3){\small $v(\emptyset,\{1,2\})$}
        \put(9.1,4.3){\small $v(\{1\},\emptyset)$}
        \put(-0.9,4.3){\small $v(\emptyset,\{1\})$}
        \put(4.1,4.3){\small $v(\emptyset,\emptyset)$}
\end{picture}
\end{center}
\caption{Ternary tuples when $n=2$}
\label{fig:tern}
\end{figure}
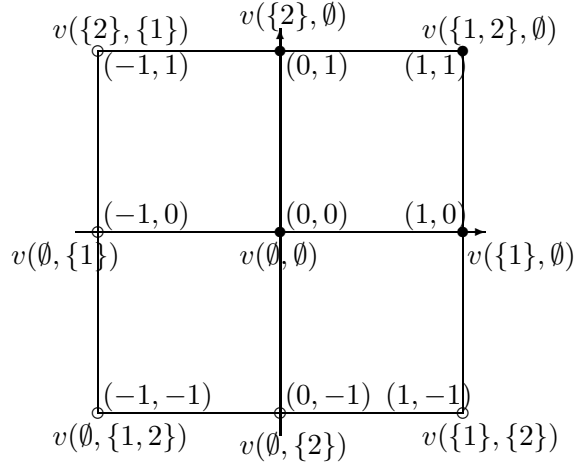

The above story for introducing interaction can be made fairly more
general. Let us first take interval $[-1,1]$ instead of $[0,1]$ for expressing
scores, and consider that for the DM, values $-1$, 0 and 1 are particular
because they express respectively total unsatisfaction, neutrality and total
satisfaction. Then we are led to consider \emph{ternary} tuples
$(1_A,-1_B,0_{(A\cup B)^c})$, whose overall score is denoted by $v(A,B)$. It is
convenient to denote by $\Q(N):=\{(A,B)\mid A,B\subseteq N, A\cap
B=\emptyset\}$. Now $v$ is defined on $\Q(N)$, and as for
capacities, it seems natural to impose $v(N,\emptyset):=1$, $v(\emptyset,
\emptyset):=0$, and $v(\emptyset,N):=-1$. Also using the dominance argument, we
should have, if $A\subseteq A'$, $v(A,B)\leq v(A',B)$ and $v(B,A)\geq
v(B,A')$. Such a $v$ is called a \emph{bi-capacity} \cite{grla02a,grla02b}. The
interaction for bi-capacities, called \emph{bi-interaction} in \cite{grla02b},
has been defined accordingly, and follows the same principle. When $n=2$, since
we have 3 particular levels $-1$, 0 and 1, the square $[-1,1]^2$ is divided
into 4 small squares and has 9 ternary tuples. In each small square, we
apply the same definition as with capacities, i.e. Eq.
(\ref{eq:int}). Hence, we have \emph{four} interaction indices to describe
interaction with $n=2$, namely (see Figure \ref{fig:tern}):
\begin{align}
I_{\{1,2\},\emptyset} & :=v(\{1,2\},\emptyset) - v(\{2\},\emptyset) -
v(\{1\},\emptyset) + v(\emptyset,\emptyset)
=:I(\{1,2\},\emptyset)\label{eq:int12}\\ 
I_{\emptyset,\{1,2\}} & :=v(\emptyset,\emptyset)
- v(\emptyset,\{1\}) - v(\emptyset,\{2\}) + v(\emptyset,\{1,2\})
=:I(\emptyset,\emptyset)\nonumber\\ 
I_{1,2} & :=v(\{1\},\emptyset) -v(\emptyset,\emptyset) -v(\{1\},\{2\})+
v(\emptyset,\{2\}) =: I(\{1\},\emptyset)\nonumber \\ 
I_{2,1} & := v(\{2\},\emptyset)
-v(\{2\},\{1\}) - v(\emptyset,\emptyset)+
v(\emptyset,\{1\})=:I(\{2\},\emptyset).\nonumber 
\end{align}
The notation $I_{A,B}$ means that criteria in $A$ are positive, while criteria in
$B$ are negative. As it will become clear later, a better notation is $I(A,B)$,
where $(A,B)$ is the ternary tuples corresponding to the upper right
corner of the square in consideration (i.e. the best possible tuple in
the square). 

Let us now take a general point of view. We consider $n$-dimensional tuples in
$X:=X_1\times\cdots\times X_n$, where it is assumed that each $X_i$ is a
partially ordered set, whose order relation is denoted by $\leq_i$. We consider
that on each dimension $X_i$, there exist \emph{reference levels}
$r^i_1,\ldots,r^i_{q_i}$, which for the problem under consideration, convey some
special meaning of interest, describing e.g. some particular situation, and that
these reference levels form a lower locally distributive lattice $(L_i,\leq_i)$.
Denoting by $L:=L_1\times\cdots\times L_n$ the product lattice with the product
order, we define a real function $v:L\longrightarrow \mathbb{R}$, assigning a
real value to any combination of reference levels on each dimension.

Let us give some instances of this general framework.
\begin{description}
\item [voting games and ternary voting games:] defining $N:=\{1,\ldots,n\}$ as
the set of voters, for each voter there exist two or three reference levels,
which are: voting in favor, voting against (case of classical voting games), and
abstention (case of ternary voting games \cite{fema97}). For classical games, we
have $L_i=\{0,1\}$, $\forall i\in N$, with level 1 corresponding to voting in
favor, so that $L=2^n$, and $v(A)=1$ if the bill is accepted when $A$ is the set
of voters voting in favor, or $v(A)=0$ if the bill is rejected. For ternary
voting games, we have $L_i=\{-1,0,1\}$, with 0 corresponding to abstention and
$-1$ to voting against. Then $L=3^n$, and it is convenient to denote an element
of $L$ by a pair $(A,B)$, where $A$ is the set of voters in favor, and $B$ the
set of voters against. As before, $v(A,B)=1$ (the bill is accepted) or 0 (the
bill is rejected). Note that here $X_i$ coincides with $L_i$, $\forall i\in N$.
\item [cooperative games and bi-cooperative games:] we replace voters by
  players. Reference levels in the case of cooperative games are 0 and 1,
  corresponding to non participation and participation in the game. Hence
  $L=2^n$, and $v(A)$ is the asset that the coalition $A$ of players will win if
  the game is played. For bi-cooperative games, $L=3^n$, and $v(A,B)$ is the
  asset that $A$ will receive when coalition $A$ plays against coalition $B$,
  the remaining players not taking part in the game. Classically, here also
  $X_i$ coincides with $L_i$, although one may consider any degree of
  participation between full participation and non participation (fuzzy games),
  which leads to $X_i=[0,1]$.
\item [multicriteria decision making:] this corresponds to the framework given
in the introduction. We have $L_i=\{0,1\}$ for all $i\in N$ if we consider only
two reference levels ``unacceptable'' and ``totally satisfying'', which leads to
capacities, and $L_i=\{-1,0,1\}$ if a neutral level is added, which leads to
bi-capacities, as explained above. Let us remark that our general framework
allows one to be much more general: one may have more than 3 levels, adding for
example intermediate levels such as ``half satisfactory'', etc., or even
introduce non comparable levels, provided the lattice structure is preserved.
For example, the level ``don't know'' may be incomparable with ``neutral'', but
smaller than ``satisfactory'' and greater than ``unsatisfactory'', thus leading
to the lattice $2^2$.
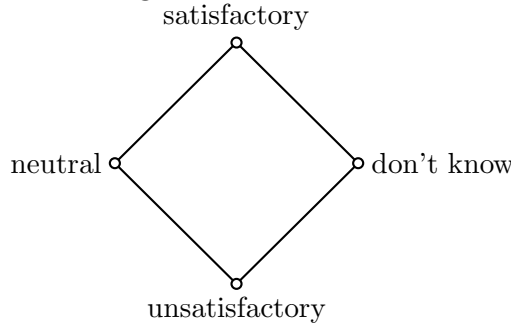
\begin{center}
\psset{unit=0.8cm}
\pspicture(0,0)(4,4)
\pspolygon(2,0)(0,2)(2,4)(4,2)
\pscircle[fillstyle=solid](2,0){0.1}
\pscircle[fillstyle=solid](0,2){0.1}
\pscircle[fillstyle=solid](2,4){0.1}
\pscircle[fillstyle=solid](4,2){0.1}
\uput[-90](2,0){\small unsatisfactory}
\uput[180](0,2){\small neutral}
\uput[90](2,4){\small satisfactory}
\uput[0](4,2){\small don't know}
\endpspicture
\end{center}
 In addition, we may consider different $L_i$ for each
criterion. The function $v$ defines the overall score given to an tuple having
various reference levels on criteria.
\item [data analysis:] the construction is the same as for multicriteria
decision making, but the meaning conveyed by the dimensions and the reference
levels can be much more general, depending on the kind of data, being for
example ``high'', ``medium'', ``low'', etc. We do not even need to have
numerical dimensions, so that ordinal data analysis can be done. The meaning of
$v(x)$ for $x\in L$ depends on the aim of the analysis. We propose three main
examples:
        \begin{itemize}
        \item evaluation of $x$. For example, $x$ is some kind of prototypical
product, and a user or consumer gives an evaluation of it, which defines $v(x)$
(subjective evaluation).
        \item classification in some category. $v(x)$ is the label of the
category, or takes value 0 or 1 (does not belong or belongs to a given
category: in this latter case we need as many functions $v$ as the number of
categories) (pattern recognition). 
\item the number of items identical or similar to $x$ in the data set (data
  mining). Suppose we have a large set $D$ of data with some distance defined on
  it.  $x\in L$ defines a particular protopyical datum. Then $v(x)$ is the
  cardinality of the set of data $x'\in D$ within a given distance of $x$, or
  $v(x)$ is the sum of the inverse distances from any $x'\in D$ to $x$. 
        \end{itemize}
\end{description}
We propose in this paper a general definition of interaction, which can be
applied to the above defined framework, and encompasses already existing
definitions of interaction for capacities and bi-capacities. The precise meaning
of interaction is governed by the meaning of the function $v$. In game theory,
it describes the synergy between players or voters, the interest to forming or
not forming certain coalitions. In multicriteria decision making, it tells which
criteria play a key role (and how), which criteria are redundant (with which
one) in the decision process. In data mining, when $v$ is a counting function as
above, the interaction has a statistical flavor close to correlation. Indeed,
since the interaction index is roughly speaking a difference of the diagonal and
anti-diagonal, a positive (resp. negative) interaction corresponds to a positive
(resp. negative) correlation. In pattern recognition, interaction is very
informative for feature selection (see an application of interaction in this
topic in \cite{gra95c}).

Clearly, the interaction is a key concept in knowledge discovery, and has a
strong descriptive power. We detail its construction and properties in the
sequel, after recalling classical results.

For simplicity, the cardinality of sets $A,B,S,\ldots$ will be denoted by the
corresponding lower case $a,b,s,\ldots$, and we will often omit braces for
singletons. We put $N:=\{1,\ldots,n\}$.
 
\section{Importance and interaction indices for $L=2^n$ and $L=3^n$} 
\label{sec:class}
We recall in this section the classical definition for $L=2^n$, (which
corresponds to capacities, or more generally set functions, pseudo-Boolean
functions \cite{haho92}), and the one for $L=3^n$ (bi-capacities, bi-cooperative
games).

Let $v:2^N\longrightarrow \mathbb{R}$, with $v(\emptyset)=0$ (game). As it will
become clear, the interaction index is a generalization of the \emph{power
index} or \emph{importance index} $\phi^v(i)$, $i\in N$, which expresses to
what extent an element $i\in N$ (attribute, dimension) has importance or power
for the problem under consideration. The general form is:
\begin{equation}
\label{eq:imp}
\phi^v(i)=\sum_{S\subseteq N\setminus i}\alpha^1_s[v(S\cup i)-v(S)],
\end{equation}
$\alpha^1_s\in\mathbb{R}$. The value of the coefficients $\alpha^1_s$ has to be
determined by additional requirements. The most important example is the
\emph{Shapley index} \cite{sha53}, where 
\begin{equation}
\label{eq:shac}
\alpha^1_s=\frac{(n-s-1)!s!}{n!},\quad s=0,\ldots,n-1,
\end{equation}
obtained by the following property: $\sum_{i=1}^n\phi^v(i)=v(N)$, expressing
a sharing of the total value among all elements, according to their
importance (\emph{efficiency} axiom). Another classical example is the Banzhaf
index \cite{ban65}, where $\alpha^1_s=\frac{1}{2^{n-1}}$, $s=0,\ldots,n-1$. 

The \emph{interaction index} \cite{gra96f} expresses the interaction among a
coalition (group) $S\subseteq N$ of elements:
\begin{equation}
\label{eq:int2}
I^v(S) = \sum_{T\subseteq N\setminus S}\alpha^s_t \Delta_Sv(T),
\end{equation}
where $\alpha_t^s\in\mathbb{R}$, and $\Delta_Sv(T)$ is the \emph{derivative} of
$v$ w.r.t. $S$ at $T$ for $S\subseteq N\setminus T$, and defined recursively as follows:
\begin{align*}
\Delta_\emptyset v(T) &:=v(T)\\
\Delta_iv(T) &:= v(T\cup i) - v(T)\\
\Delta_Sv(T) &:= \Delta_i(\Delta_{S\setminus i}v(T)), \quad 
|S|>1. 
\end{align*}
Observe that $I^v(\{i\})\equiv \phi^v(i)$, hence an interaction index is a
generalization of an importance index. It is possible to define recursively the
interaction index from the importance index \cite{grro97a}. Then, choosing a
particular importance index (hence the coefficients $\alpha^1_s$) defines
uniquely the coefficients $\alpha^s_t$. Let us introduce some notations,
borrowed from game theory. The \emph{restricted game} $v^{N\setminus K}$ is the
game $v$ restricted to elements (players) in $N\setminus K$, hence
$v^{N\setminus K}(S)=v(S)$ for any $S\subseteq N\setminus K$, and is not defined
outside. The \emph{reduced game} $v^{[K]}$ is the game where all elements in $K$
are considered as a single element denoted by $[K]$, i.e. the set of elements is
then $N_{[K]}:=(N\setminus K)\cup \{[K]\}$. The
reduced game is defined by, for any $S\subseteq N\setminus K$:
\begin{align*}
v_{[K]}(S) & =  v(S)\\
v_{[K]}(S\cup\{[K]\}) & =  v(S\cup K).
\end{align*}
The recursion axiom writes
\begin{equation}
\label{eq:recur}
I^v(S) = I^{v^{[S]}}([S]) - \sum_{K\subseteq S,
K\neq\emptyset,S}I^{v^{N\setminus K}}(S\setminus K).
\end{equation}
Its meaning is simple when $|S|=2$. Indeed, the formula can be written as
\[
I^{v^{[i,j]}}([i,j])= I^{v^{N\setminus i}}(j) + I^{v^{N\setminus j}}(i) + I^v(i,j).
\]
It means that the importance of elements (e.g. players) $i,j$ taken together is
the sum of individual importances when the other is absent, and the interaction
they have between them. Hence a positive interaction means that the
\emph{overall} importance of $i,j$ is greater than the sum of their respective
\emph{marginal} importances (see \cite{grro97a} for another equivalent axiom).

This axiom leads to the following formula for $\alpha_s^t(n)$, the argument
indicating the number of players in the game
\begin{equation}
\label{eq:coef}
\alpha_s^t(n)=\alpha_s^1(n-t+1),\quad \forall s=0,\ldots,n-t,\quad \forall t=1,\ldots,n-1. 
\end{equation}
When $\phi^v$ is the Shapley index, we obtain the
\emph{Shapley interaction index}, whose coefficients are, using (\ref{eq:coef}):
\[
\alpha^s_t:=\frac{(n-s-t)!t!}{(n-s+1)!}.
\]

\medskip

We have generalized the above notions to the case of bi-capacities and
bi-cooperative games \cite{grla02b,grla03d}, and given an axiomatization
\cite{grla03d,lagr03a}. As explained in Section \ref{sec:intro}, we have to
consider all combinations between positive and negative parts of the $X_i$'s
(see Eq. (\ref{eq:int12})), and following the notation introduced there, we
denote by $I_{S,T}$, $(S,T)\in\Q(N)$, the interaction among elements when $S$ is
the set of positive elements, and $T$ is the set of negative elements.  The
Shapley index divides into two indices $I_{\{i,\emptyset\}}$ and
$I_{\{\emptyset,i\}}$, defined by:
\begin{align}
I_{\{i,\emptyset\}} & := \sum_{S\subseteq N\setminus
  i}\frac{(n-s-1)!s!}{n!}\Delta_{i,\emptyset}v(S,N\setminus(S\cup
  i))\label{eq:imp1} \\
I_{\{\emptyset,i\}} & := \sum_{S\subseteq N\setminus
  i}\frac{(n-s-1)!s!}{n!}\Delta_{\emptyset,i}v(S,N\setminus S) \label{eq:imp2}
\end{align}
where the derivatives are defined by:
\begin{align*}
\Delta_{i,\emptyset}v(S,T) & := v(S\cup i,T) - v(S,T),\quad
(S,T)\in\Q(N\setminus i)\\ 
\Delta_{\emptyset,i}v(S,T) & := v(S,T\setminus i) - v(S,T), \quad (S,T)\in\Q(N),
S\not\ni i, T\ni i.
\end{align*}
$\Delta_{i,\emptyset}v(S,T)$ is the contribution of element $i$ when it acts as
a positive element, while $\Delta_{\emptyset,i}v(S,T)$ is the (negative)
contribution of $i$ when acting as a negative element. Hence the two above
Shapley values are average contributions of an element when it acts as a positive
or as a negative element.

The coefficients are obtained through an efficiency axiom which reads:
\[
\sum_{i\in N}\Big[I(i,\emptyset) + I(\emptyset,i)\Big] = v(N,\emptyset) - v(\emptyset,N). 
\]
As above, the derivative $\Delta_{S,T}$ can be defined recursively from these
equations, and the definition of the Shapley interaction index is:
\[
I_{S,T} := \sum_{K\subseteq N\setminus (S\cup
T)}\frac{(n-s-t-k)!k!}{(n-s-t+1)!}\Delta_{S,T}v(K,N\setminus(K\cup S)).
\]

\section{Mathematical background and general framework for interaction}
\label{sec:math}
We try now to have a general view of previous definitions, thanks to results
from lattice theory. We first introduce necessary definitions (see
e.g. \cite{bir67,dapr90,gratz98}).

\begin{figure}[htb]
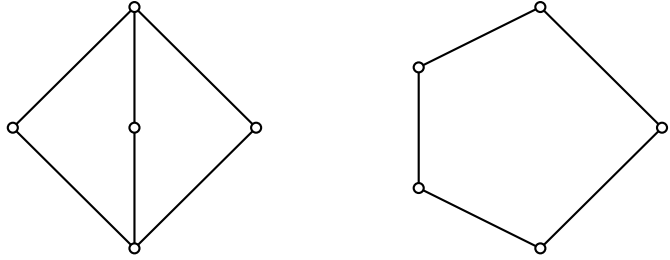

\begin{center}
\psset{unit=0.8cm}
\pspicture(0,0)(4,4)
\pspolygon(2,0)(0,2)(2,4)(4,2)
\psline(2,0)(2,4)
\pscircle[fillstyle=solid](2,0){0.1}
\pscircle[fillstyle=solid](0,2){0.1}
\pscircle[fillstyle=solid](2,4){0.1}
\pscircle[fillstyle=solid](4,2){0.1}
\pscircle[fillstyle=solid](2,2){0.1}
\endpspicture
\hspace*{2cm}
\pspicture(0,0)(4,4)
\pspolygon(2,0)(0,1)(0,3)(2,4)(4,2)
\pscircle[fillstyle=solid](2,0){0.1}
\pscircle[fillstyle=solid](0,1){0.1}
\pscircle[fillstyle=solid](0,3){0.1}
\pscircle[fillstyle=solid](2,4){0.1}
\pscircle[fillstyle=solid](4,2){0.1}
\endpspicture
\end{center}
\caption{The lattices $M_3$ (left) and $N_5$ (right)}
\label{fig:M3}
\end{figure}

Let $(L,\leq)$ be a lattice, we denote as usual by $\vee,\wedge,\top,\bot$
supremum, infimum, top and bottom (if they exist). If $x$ and $y$ in $L$ are
incomparable, we write $x||y$. $Q\subseteq L$ is a \emph{downset} of $L$ if
$x\in Q$ and $y\leq x$ imply $y\in Q$. For any $x\in L$, the \emph{principal
  ideal} $\downarrow x$ is defined as $\downarrow x:=\{y\in L\mid y\leq x\}$
(downset generated by $x$). For $x,y\in L$, we say that $x$ \emph{covers} $y$
(or $y$ is a \emph{predecessor} of $x$), denoted by $x\succ y$, if there is no
$z\in L, z\neq x,y$ such that $x\geq z\geq y$. $(L,\leq)$ is \emph{lower
  semi-modular} (resp. \emph{upper semi-modular}) if for all $x,y\in L$, $x\vee
y\succ x$ and $x\vee y\succ y$ imply $x\succ x\wedge y$ and $y\succ x\wedge y$
(resp. $x\succ x\wedge y$ and $y\succ x\wedge y$ imply $x\vee y\succ x$ and
$x\vee y\succ y$). A lattice being upper and lower semi-modular is called
\emph{modular}. A lattice is modular iff it does not contain $N_5$ as a
sublattice (see Fig. \ref{fig:M3}). A lattice is \emph{distributive} when
$\vee,\wedge$ satisfy the distributivity law, and it is \emph{complemented} when
each $x\in L$ has a (unique) complement $x'$, i.e. satisfying $x\vee x'=\top$
and $x\wedge x'=\bot$. A modular lattice is distributive iff it does not contain
$M_3$ as a sublattice (see Fig. \ref{fig:M3}). A lattice is \emph{linear} if it
is totally ordered. A lattice is said to be
\emph{Boolean} if it has a top and bottom element, is distributive and
complemented. When $L$ is finite, it is Boolean iff it is isomorphic to the
lattice $2^n$ for some $n$.

$(L,\leq)$ is said to be \emph{lower locally distributive} if it is lower
semi-modular, and it does not contain a sublattice isomorphic to $M_3$. 
Equivalently, it is lower locally distributive if for any $x\in L$, the
interval $[\bigwedge_{y\prec x}y,x]$ is a Boolean lattice (see
\cite{mon90} for a survey).  

An element $i\in L$ is \emph{join-irreducible} if it cannot be written as a
supremum over other elements of $L$ and it is not the bottom element. When $L$
is finite, this is equivalent to $i$ covers only one element. Let us call
$\mathcal{J}(L)$ the set of all join-irreducible elements of $L$.

In a finite distributive lattice, any element $y\in L$ can be decomposed in
terms of join-irreducible elements. The fundamental result due to Birkhoff is
the following.
\begin{theorem}
\label{th:birk}
Let $L$ be a finite distributive lattice. Then the map $\eta:L\longrightarrow
\mathcal{O}(\mathcal{J}(L))$, where $\mathcal{O}(\mathcal{J})$ is the set of all
downsets of $\mathcal{J}$, defined by
\[
\eta(x) :=\{i\in\mathcal{J}(L)\mid i\leq x\}=\mathcal{J}(L)\cap \downarrow x
\]
is an isomorphism of $L$ onto $\mathcal{O}(\mathcal{J}(L))$.
\end{theorem}
We call $\eta(x)$ the \emph{normal decomposition} of $x$, we have 
\[
x=\bigvee \eta(x).
\] 
The isomorphism says that $x\leq y$ iff $\eta(x)\subseteq \eta(y)$, hence
$\eta(x\vee y) =\eta(x)\cup\eta(y)$ and so on.

The decomposition of some $x$ in $L$ in term of supremum of join-irreducible
elements is unique up to the fact that it may happen that some join-irreducible
elements in $\eta(x)$ are comparable. Hence, if $i\leq j$ and $j$ is in a
decomposition of $x$, then we may delete $i$ from the decomposition of $x$. We
call \emph{minimal decomposition} the (unique) decomposition of minimal
cardinality, denoted by $\eta^*(x)$. \emph{Atoms} are join-irreducible elements
covering $\bot$. A lattice is \emph{atomistic} if all join-irreducible elements
are atoms. A finite distributive atomistic lattice is Boolean.

As shown by Dilworth \cite{dil40}, any $x\in L$ has a unique join-irreducible
minimal decomposition iff it is lower locally distributive.

A useful result is the following
\begin{equation}
\label{eq:down}
\downarrow x = \{y\mid \eta(y) = \bigcup_{j\in K}j,\quad K\subseteq \eta(x)\}.
\end{equation}

When $L=2^n$, join-irreducible elements are simply atoms (i.e. singletons of
$N$). When $L=3^n$, join-irreducible elements are $(i,i^c)$ and
$(\emptyset,i^c)$, $\forall i\in N$.

Let $(L,\leq)$ be a locally finite partially ordered set, and a function
$g:L\longrightarrow \mathbb{R}$. Consider the following equation
\begin{equation}
\label{eq:mob1}
g(x) = \sum_{y\leq x} f(y).
\end{equation}
There is a unique solution $f:L\longrightarrow \mathbb{R}$ to this equation,
called the \emph{M\"obius transform} of $g$ (see Rota \cite{rot64}). Note that
in a sense, $f$ could be considered as the derivative of $g$.

\medskip

As said in the introduction, our general framework for the definition of
interaction will be to consider finite lower locally distributive lattices
$L_1,\ldots,L_n$, with top and bottom of $L_i$ denoted by $\top_i,\bot_i$,
$i=1,\ldots,n$, and the product lattice $L:=L_1\times\cdots\times L_n$ with the
product order. Sometimes, we will need in addition that the $L_k$'s are modular
(hence they are distributive). We set $N:=\{1,\ldots,n\}$. A \emph{vertex} of
$L$ is an element $x=(x_1,\ldots,x_n)$ of $L$ where $x_i$ is either $\top_i$ or
$\bot_i$, for $i=1,\ldots,n$. We denote by $\Gamma(L)$ the set of vertices of
$L$. Note that if $L$ is a Boolean lattice, then $L=\Gamma(L)$. For $\Q(N)$,
vertices are of the form $(A,A^c)$, $A\subseteq N$.

Since $L_i$ is finite and lower locally distributive, it can be represented by
join-irreducible elements. Then join-irreducible elements of $L$ are simply
of the form 
\[
i=(\bot_1,\ldots,\bot_{j-1},i_0,\bot_{j+1},\ldots,\bot_n),
\]
for some $j\in\{1,\ldots,n\}$ and some $i_0\in \mathcal{J}(L_j)$. Hence,
there are $\sum_{j=1}^n |\mathcal{J}(L_j)|$ join-irreducible elements in $L$.

\section{Derivative of a function over a lattice}
Let $(L,\leq)$ be a finite lower locally distributive lattice, and
$f:L\longrightarrow \mathbb{R}$ a real-valued function on it. 
\begin{definition}
Let $i\in \mathcal{J}(L)$. The \emph{derivative} of $f$
w.r.t. $i$ at point $x\in L$ is given by:
\[
\Delta_if(x) := f(x\vee i) - f(x).
\]
\end{definition}
Note that $\Delta_if(x) = 0 $ if $i\leq x$. We say that the derivative
$\Delta_if(x)$ is \emph{Boolean} if $[x,x\vee i]$ is the Boolean lattice $2^1$,
otherwise said $x\vee i\succ x$. Differentiating two times w.r.t two
join-irreducible elements $i,j$ such that $i||j$ ($i$ and $j$ are incomparable)
leads to:
\[
\Delta_i(\Delta_j f(x)) = \Delta_j(\Delta_i f(x)) = f(x\vee
i\vee j)- f(x\vee i) - f(x\vee j) + f(x). 
\]
We call this quantity the \emph{second derivative} w.r.t $i,j$ or the derivative
w.r.t $i\vee j$, denoted by $\Delta_{i\vee j}f(x)$. Note that allowing $i\leq j$
leads to $\Delta_{i\vee j}f(x)= -\Delta_if(x)$.

Using the minimal decomposition, the derivative w.r.t any element
$y$ can be defined.
\begin{definition}
Let $x,y\in L$, and $y=\vee_{k=1}^n i_k$ be the minimal decomposition of $y$
into join-irreducible elements. Then the derivative of $f$ w.r.t $y$ at point
$x$ is given by:
\[
\Delta_y f(x) = \Delta_{i_1}(\Delta_{i_2}(\cdots\Delta_{i_n}f(x)\cdots)).
\] 
\end{definition}
The derivative is \emph{Boolean} if $[x,x\vee y]$ is the Boolean lattice
$2^n$. The derivative is 0 if for some $k$, $i_k\leq x$. The following lemma
gives practical equivalent conditions.
\begin{lemma}
\label{lem:incr}
Let $x,y\in L$. 
\begin{itemize}
\item [(i)] The
derivative $\Delta_yf(x)$ is 0 whenever $\eta(x)\cap \eta^*(y)\neq \emptyset$. 
\item [(ii)] The derivative $\Delta_yf(x)$ is Boolean iff $\eta(x\vee y) = \eta(x)\cup\bigcup\eta^*(y)$.
\end{itemize}
\end{lemma}
\begin{proof}
(i) Let $k\in \eta(x)\cap \eta^*(y)$. Since $k\in \eta(x)$, all join-irreducible elements below $k$
are also in $\eta(x)$, hence   $\eta(k)\subseteq\eta(x)$. By
Th. \ref{th:birk}, this is equivalent to $k\leq  x$, which in turn
implies that the derivative is 0 since $k\in \eta^*(y)$.

\noindent
(ii) Let us consider first $y=i\in\mathcal{J}(L)$, and suppose $\Delta_if(x)$
is Boolean. Since $x\vee i\succ x$, by
isomorphism, we have $\eta(x\vee i)\succ\eta(x)$, which means that there exists
some $k\in \mathcal{J}(L)$ such that $\eta(x\vee i)=\eta(x)\cup\{k\}$. Since
$\eta(x\vee i)=\eta(x)\cup\eta(i)$, $k\in\eta(i)$, and all other $j\in\eta(i)$
belong also to $\eta(x)$. Hence $k=i=\eta^*(i)$ since
$\eta(i)=\{j\in\mathcal{J}(L)\mid j\leq i\}$. The converse is clear. Applying
recursively this result proves (ii).
\end{proof}

As a consequence of (ii), the lattice $[x,x\vee y]$ is isomorphic to $(\mathcal{P}(\eta^*(y)),\subseteq)$.

We express the derivative in terms of the M\"obius transform of $f$. 
\begin{proposition}
Let $i$ be a join-irreducible element such that $\Delta_if(x)$ is
Boolean. We denote by $m$ the M\"obius transform of $f$. Then
\[
\Delta_i f(x) = \sum_{y\in [i,x\vee i]} m(y).
\]
\end{proposition}
\begin{proof}
We have:
\[
\Delta_i f(x) = \sum_{y\leq x\vee i}m(y) - \sum_{y\leq x}m(y) =
\sum_{y\in\downarrow(x\vee i)\setminus \downarrow x}m(y),
\]
since $\downarrow x\subset\downarrow (x\vee i)$.  Using Lemma \ref{lem:incr}
(ii), we have $\eta(x\vee i)=\eta(x)\cup\{i\}$. Applying (\ref{eq:down}), we
get
\[
\downarrow(x\vee i)\setminus \downarrow x = \{y\mid \eta(y)=\bigcup_{j\in K}j\cup \{i\},\quad
K\subseteq \eta(x)\}=[i,x\vee i]
\]
since we get $i$ for $K=\emptyset$, and $x\vee i$ for $K=\eta(x)$, and the set
is clearly an interval.
\end{proof}

Based on this, we can show the general result:
\begin{theorem}
\label{th:dermob}
Let $x,y\in L$, such that $\Delta_yf(x)$ is Boolean. Then
\[
\Delta_yf(x) = \sum_{z\in[y,x\vee y]}m(z).
\]  
\end{theorem}
\begin{proof}
We proceed by recurrence on $|\eta^*(y)|$. The result is already shown for
$|\eta^*(y)|=1$. Let us suppose it holds for some $y$, and consider $y'=y\vee
i$, with $i\not\in \eta(y)$ and $\Delta_{y'}f(x)$ being Boolean. We have:
\begin{align*}
\Delta_{y'}f(x) & = \Delta_i(\Delta_y f(x))\\
        & = \Delta_y f(x\vee i) - \Delta_y f(x)\\
        & = \sum_{z\in[y,x\vee y\vee i]}m(z) - \sum_{z\in[y,x\vee y]}m(z)\\
        & = \sum_{z\in [y,x\vee y\vee i]\setminus [y,x\vee y]}m(z).
\end{align*}
Since $[y,x\vee y]=\{z\mid \eta(z)=\eta(y)\cup \bigcup_{j\in J}j, J\subseteq
\eta(x)\}$ and $[y,x\vee y\vee i] = \{z\mid \eta(z) = \eta(y)\cup \bigcup_{j\in
J}j,J\subseteq \eta(x)\cup \{i\}\}$ we get
\[
[y,x\vee y\vee i]\setminus[y,x\vee y] = \{z\mid \eta(z) =
\eta(y)\cup\bigcup_{j\in J}j\cup\{i\}, J\subseteq \eta(x)\} = [y',y'\vee x],
\]
the desired result.
\end{proof}

The close link between our derivative and M\"obius transform is not surprising
since the M\"obius transform has already a meaning of derivative. 

Let us apply these results to the case of usual capacities and
bi-capacities. It suffices to check if formulas coincide for join-irreducible
elements. For capacities, we have for any $i\in N$, $\Delta_i v(A):=v(A\cup i)
-v(A)$, so that we recover the definition above. Note that this coincides with
the notion of derivative for pseudo-Boolean functions \cite{haho92}. For
bi-capacities, we have
\begin{align*}
\Delta_{(i,i^c)}v(A,B) & = v(A\cup i, B) - v(A,B) = \Delta_{i,\emptyset}
v(A,B)\\ 
\Delta_{(\emptyset,i^c)}v(A,B) & = v(A, B\setminus i) - v(A,B) = 
\Delta_{\emptyset,i} v(A,B),
\end{align*}
which again coincides with the definition given above, although notation
differs.

\section{Interaction: the general case}
\label{sec:intgen}
As seen in Section \ref{sec:class}, the definition of the derivative is the key
concept for the interaction index. Using our general definition of derivative
with new notation, let us express the interaction when $L=3^n$ using the
notation $\Delta_{(S,T)}$. Imposing the same argument to $I$ and $\Delta$, we
are led to:
\begin{equation}
\label{eq:intst}
I^v(S,T) = \sum_{K\subseteq
T}\frac{(t-k)!k!}{(t+1)!}\Delta_{(S,T)}v(K,N\setminus(K\cup S)),
\end{equation}
with the correspondence $I^v_{S,T} = I^v(S,(S\cup T)^c)$. Observe that these
two notations precisely correspond to those introduced in
Eqs. (\ref{eq:int12}).

We remark that the derivative in the above expression is taken over some
vertices of $\Q(N\setminus S)$. Also, the importance index corresponds to
derivatives w.r.t. join-irreducible elements. Based on these observations, we
are now in position to propose a definition using our general framework
(see Section \ref{sec:math}). Roughly speaking, the interaction index
w.r.t. $x\in L$ is a weighted average of the derivative w.r.t. $x$, taken at
vertices of $L$ ``not related'' to $x$. The weights can be determined
recursively from the cases where $x$ is a join-irreducible element, and the
coefficients for these cases are determined by some normalization condition
(e.g. efficiency-like condition in the case of the Shapley index).

\subsection{Definition of interaction}
We begin by defining the importance index, i.e. interaction index w.r.t. a
join-irreducible element.   
\begin{definition}
\label{def:inter}
Let $i=(\bot_1,\ldots,\bot_{j-1},i_0,\bot_{j+1},\ldots,\bot_n)$ be a
join-irreducible element of $L$. The \emph{interaction w.r.t. $i$} of $v$ is
any function of the form
\begin{equation}
\label{eq:inter}
I(i) := \sum_{x\in\Gamma(\prod_{k=1}^{j-1}L_k)\times\{\underline{i_0}\}\times\Gamma(\prod_{k=j+1}^nL_k)}\alpha^1_{h(x)}\Delta_i
v(x),
\end{equation}
where $\underline{i_0}$ is the (unique) predecessor of $i_0$ in $L_j$, $h(x)$
is the number of components of $x$ equal to $\top_l$, $l=1,\ldots,n$, and
$\alpha^1_k\in\mathbb{R}$ for any integer $k$.
\end{definition}
Observe that the constants $\alpha^1_{h(x)}$ do not depend on $i$.  Also, the
derivative is Boolean.

Let us show that this definition encompasses the case of capacities and
bi-capacities. For capacities, $L_k=\{0,1\}$ for all $k$, with 1 as unique
join-irreducible element, join-irreducible elements of $L=2^n$ are
singletons, all elements in $L$ are vertices, and $h(x)$ is the
cardinality of sets. Thus we get for a singleton $j\in N$:
\[
I(j) = \sum_{A\subseteq N\setminus j}\alpha^1_{|A|}[v(A\cup j)-v(A)]
\]
as desired. For bi-capacities, $L_k=\{-1,0,1\}$ for all $k$, with
$\mathcal{J}(L_k)=\{0,1\}$. The height function is $h(A,B) = |A|$. Let us
consider first the case where the join-irreducible element in $L=3^n$ is
$(j,j^c)$, or in vector form $(-1,\ldots,-1,1,-1,\ldots,-1)$, where 1 is at the
$j$th place. Then $\Gamma(3^{j-1})\times\{0\}\times \Gamma(3^{n-j})$ corresponds
to vertices of $\Q(N\setminus j)$. Thus we obtain
\[
I(j,j^c)  =\sum_{A\subseteq N\setminus
j}\alpha^1_{|A|}\Delta_{(j,j^c)}v(A,N\setminus(A\cup j))
\]
which has the required form. Let us examine now the case of $(\emptyset,j^c)$,
which is, in vector form, $(-1,\ldots,-1,0,-1,\ldots,-1)$. This time
$\Gamma(3^{j-1})\times\{-1\}\times \Gamma(3^{n-j})$ is $\Gamma(L)$, after
removal of vertices $(A,A^c)$ with $j\in A$.  In summary, we obtain:
\[
I(\emptyset,j^c) = \sum_{A\subseteq N\setminus
j}\alpha^1_{|A|}\Delta_{(\emptyset,j^c)} v(A, N\setminus A)
\]
which has again the required form. 

Let us generalize Def. \ref{def:inter} to a class of elements of $L$
denoted by $\tilde{L}$ and defined as follows: $\tilde{L}:=\bigcup_{K\subseteq
  N}\tilde{L}_K$, with 
\[
\tilde{L}_K:= \{x\in L\mid\forall k\in K,  \exists !\
\underline{i_k}\in L_k \text{ such that }\forall i\in \eta^*(x_k), i\succ
\underline{i_k}, \text{ and }x_k=\bot_k\text{ if }k\in N\setminus K\}
\]  
In words, it is the set of elements whose coordinates are either bottom or such
that the minimal decomposition covers a unique element. Observe that for the
case where $L_k$ is a linear lattice or an atomistic one (i.e. practical cases
of interest), $\tilde{L}=L$.
\begin{definition}
\label{def:intg}
Let $K\subseteq N$, $x\in \tilde{L}_K$, and denote as above by
$\underline{i_k}$, for all $k\in K$, the element covered by all
$i\in\eta^*(x_k)$.  The \emph{interaction w.r.t. $x$} of $v$ is any function of
the form
\begin{equation}
\label{eq:genint}
I(x) := \sum_{y\mid y_k=\top_k \text{ or } \bot_k \text{ if } k\not\in K,
y_k =\underline{i_k} \text{ else}}\alpha^{|K|}_{h(y)}\Delta_x v(y)
\end{equation}
where $h(y)$ is the number of components of $y$ equal to $\top_l$,
$l=1,\ldots,n$.
\end{definition}

The derivative is Boolean if in addition the $L_k$'s are modular
(and hence distributive), by application of the following Lemma.
\begin{lemma}
  \label{lem:boo} If $L_k$ is distributive, $k=1,\ldots,n$, then for any
  $K\subseteq N$, any $x\in \tilde{L}_K$, $\Delta_xv(y)$ is Boolean for any $y$
  such that $y_k=\top_k$ or $\bot_k$, $k\not\in K$, and $y_k=\underline{i_k}$,
  where $\underline{i_k}$ is the element covered by all $i\in\eta^*(x_k)$.
\end{lemma}
\begin{proof}
We have to prove that $[y,x\vee y]$ is isomorphic to $2^{|\eta^*(x)|}$, with $y$
defined as above. It suffices to prove that $[y_k,x_k\vee y_k]$ is isomorphic to
$2^{|\eta^*(x_k)|}$ for each coordinate $k$. If $k\not\in K$, then $[y_k,x_k\vee
y_k]=\{y_k\}\cong 2^0$. If $k\in K$, then $y_k$ is covered by all $i$ in
$\eta^*(x_k)$. Hence $[y_k,x_k\vee
y_k]=[\underline{i_k},x_k]$, which is atomistic. Since it is also distributive,
it is Boolean and isomorphic to $2^{|\eta^*(x_k)|}$. 
\end{proof}

\subsection{Expression with the M\"obius transform and efficiency}
Let us express $I(x)$ w.r.t the M\"obius transform. First we recall the result
for bi-capacities, which writes \cite{grla02b,grla03d}:
\[
I(S,T) = \sum_{(S',T')\in[(S,T),(S\cup T,\emptyset)]}\frac{1}{t-t'+1}m(S',T').
\]
We have the following general result.
\begin{theorem}
\label{th:intmob}
Let $K\subseteq N$, and assume distributivity holds for every $L_k$, $k\in K$.
The expression of the interaction index for $x\in \tilde{L}_K$ in terms of the
M\"obius transform is given by:
\[
I(x) = \sum_{z\in[x, \check{x}]}\beta^{|K|}_{k(z)}m(z),
\]
with $\check{x}_k:=(\top_k)$ for $k\not\in K$, $\check{x}_k=x_k$ else, and
$k(z)$ is the number of coordinates of $z$ not equal to $\bot_l$,
$l=1,\ldots,n$. Moreover, the real constants $\beta^{|K|}_{k(z)}$ are
related to the $\alpha^{|K|}_{h(x)}$'s by:
\begin{equation}
\label{eq:beta}
\beta^{|K|}_{k(z)} =
\sum_{l=0}^{n-k(z)}\binom{n-k(z)}{l}\alpha^{|K|}_{ (k(z)-|K|+l)}
\end{equation}
\end{theorem}
\begin{proof}
  Since the derivative is Boolean by Lemma \ref{lem:boo}, we can apply Th.
  \ref{th:dermob}, and we get:
\begin{equation}
\label{eq:11}
I(x) = \sum_{y\mid y_k=\top_k \text{ or } \bot_k \text{ if } k\not\in K,
y_k =\underline{i_k} \text{ else}}\alpha^{|K|}_{h(y)}\sum_{z\in[x,y\vee
x]}m(z). 
\end{equation}
Then for any $y$ such that $y_k=\top_k$ or $\bot_k$ if $k\not\in K$, and $y_k
=\underline{i_k}$ else, $(y\vee x)_k = x_k$ when $k\in K$, other coordinates
being $\top_k$ or $\bot_k$, in any combination. Hence for all possible such
$y$, $z$ takes any value in
$[x,(\top_{N\setminus K},x_K)]$, where $(\top_{N\setminus K},x_K)$ has
coordinate $\top_k$ when $k\not\in K$, and $x_k$ else. Denoting
by $\check{x}$ the right bound of this interval, we get
\[
I(x) = \sum_{z\in[x,\check{x}]}\beta_zm(z).
\]
It remains to express $\beta_z$ in terms of $\alpha^{|K|}_{h(x)}$. Let us take
a fixed $z\in [x,\check{x}]$ and examine for which $y$'s in (\ref{eq:11}) it
belongs to $[x,y\vee x]$. Note that $z_k=x_k$ for all $k\in K$. Since $y_l$,
$l\not\in K$ is either $\bot_l$ or $\top_l$, we must have $y_l=\top_l$ whenever
$z_l\neq\bot_l$, the other coordinates not in $K$ being free.  The result is
then:
\[
\beta_z = \sum_{y\mid y_l=\top_l\text{ if } z_l\neq\bot_l, l\not\in K}
\alpha^{|K|}_{h(y)}. 
\]
Denoting by $k(z)$ the number of coordinates not equal to $\bot_l$, we get
\[
\beta_z = \sum_{l=0}^{n-k(z)}\binom{n-k(z)}{l}\alpha^{|K|}_{k(z)-|K|+l}
\]
Remarking that $\beta_z$ depends only on $k(z)$ and $|K|$, we get the
desired result.
\end{proof}

Let us check if we recover the coefficients for bi-capacities and Shapley
index.  For $(S,T)=(i,i^c)$ and $(\emptyset,i^c)$, we have
$\beta_{(S',T')}=\frac{1}{n-t'}$. We apply (\ref{eq:beta}), noting that
$(S',T')$ has $n-t'$ coordinates different from bottom:
\begin{align*}
\beta_{(S',T')} & = \sum_{l=0}^{t'}\binom{t'}{l}\alpha^1_{n-t'-1+l}\\
 & = \sum_{l=0}^{t'}\binom{t'}{l}\frac{(n-t'-1+l)!(t'-l)!}{n!}\\
& = \sum_{l=0}^{t'}\frac{t'!(n-t'-1+l)!}{l!n!}.
\end{align*} 
In \cite{grla03d}, the following combinatorial result was shown:
\[
\sum_{i=0}^k \frac{(n-i-1)!k!}{n!(k-i)!} = \frac{1}{n-k}.
\]
Applying the above formula with $i=t'-l$, we get the
desired result.

\medskip

It is possible to find easily the $\beta_{k(z)}^{1}$ coefficients if we
consider a normalization condition as for the Shapley index. Let us define
\emph{efficiency} as
\begin{equation}
\label{eq:shap}
\sum_{i\in \mathcal{J}(L)}I(i)=v(\top) -v(\bot),
\end{equation}
and call \emph{Shapley interaction index} the resulting interaction index.
Applying Th. \ref{th:intmob}, we get:
\[
\sum_{i\in \mathcal{J}(L)}I(i)= \sum_{i\in \mathcal{J}(L)}\sum_{z\in
[i,\check{\imath}]}\beta^{1}_{k(z)}m(z).
\]
Let us take $m$ such as it is non zero only for a given $z\in L$, say $z_0$,
such that for all coordinates $z_l$ different from bottom, we have
$z_l\in\mathcal{J}(L_l)$. Since $\sum_{x\in L}m(x) = v(\top)$, we have
necessarily $m(z_0)=v(\top)-v(\bot)$. Observe that $z_0$ belongs to all
intervals $[i,\check{\imath}]$ such that $z_0\geq i$ and $z_0\leq
\check{\imath}$.  Recalling that
$i=(\bot_1,\ldots,\bot_{j-1},i_0,\bot_{j+1},\ldots,\bot_n)$ and $\check{\imath}
= (\top_1,\ldots,\top_{j-1},i_0,\top_{j+1},\ldots,\top_n)$, if $z$ has only
coordinate $z_l\neq\bot_l$, then only $i$ such that $i_l=z_l$ is suitable. More
generally, if $z$ has only $k$ coordinates different from bottom, then we have
only $k$ choices for $i$. Hence, for such $z$
\[
\beta^{1}_{k(z)} = \frac{v(\top)
-v(\bot)}{k(z)[v(\top)-v(\bot)]}=\frac{1}{k(z)}.
\]

Let us apply this to the Shapley index for bi-capacities. We get:
\[
\beta^{1}_{(S',T')}=\frac{1}{n-t'}.
\]
Suppose the $\beta^{1}_{k(z)}$'s are determined by some rule, as above. Since
$k(z)$ takes values in $\{1,\ldots,n\}$, there are $n$ coefficients
$\beta^{1}_{k(z)}$, while for $\alpha^1_{h(x)}$,\linebreak
$h(x)\in\{0,\ldots,n-1\}$, so that there are also $n$ coefficients. Th.
\ref{th:intmob} tells us that $\alpha^1_0,\ldots,\alpha^1_{n-1}$ can be computed
from $\beta^{1}_1,\ldots,\beta^{1}_n$ by solving the triangular linear system
(\ref{eq:beta}). Since there is no 0 on the diagonal of the matrix, there is
always a unique solution to this system.

Applying this observation to the Shapley interaction index, we get the following
result. 
\begin{theorem}
When $L_k$ is distributive, for all $k=1,\ldots,n$, the coefficients
$\alpha^1_0,\ldots,\alpha^1_{n-1}$ of the  Shapley interaction index $I(i)$,
$i\in \mathcal{J}(L)$ (i.e. satisfying Def. \ref{def:inter} and (\ref{eq:shap}))
are given by:
\[
\alpha^1_k = \frac{(n-1-k)!k!}{n!}.
\]
\end{theorem}

\subsection{The recursion axiom for the linear case}
Let us generalize the recursion axiom (\ref{eq:recur}) to compute $I(x)$, with
the following additional restriction: all $L_k$'s are linear lattices. Hence,
all previous results apply.  Also, all derivatives involved
are Boolean.

Let $J\subseteq N$, and consider $x$ such that $x_k=\bot_k$ if $k\not\in J$, and
$x_k=i_k$ else, for some $i_k\in \mathcal{J}(L_k)$. We denote as before by
$\underline{i_k}$ the unique predecessor of $i_k$. We introduce additional
notations. For any $K\subseteq J, K\neq \emptyset,J$, the function $v$
\emph{restricted to} $\prod_{k\in N\setminus K}L_k$ is denoted by $v^{N\setminus
  K}_x$ and defined by:
\[
v^{N\setminus K}_x(y) := v(y'), \text{ with } y'_k:=\begin{cases}
                                        \underline{i_k}, & \text{ if }
                                        k\in K\\
                                        y_k, & \text{else}
                                        \end{cases},\quad \forall y\in
                                        \prod_{k\in N\setminus K}L_k. 
\] 
The function $v$ \emph{reduced to} $x$ is a function $v^{[x]}$ defined on
$\prod_{k\in N\setminus J}L_k \times \{\bot_{[x]},\top_{[x]}\}$ by:
\[
v^{[x]}(y) : = v(\phi_{[x]}(y)), \quad\forall y\in \prod_{k\in N\setminus J}L_k
\times \{\bot_{[x]},\top_{[x]}\}, 
\]
and $\phi_{[x]}:\prod_{k\in N\setminus J}L_k \times
\{\bot_{[x]},\top_{[x]}\}\longrightarrow L$ is defined by
\[
\phi_{[x]}(y) := y', \text{ with }y'_k:=\begin{cases}
                i_k, & \text{ if } k\in J \text{ and } y_{[x]}=\top_{[x]} \\
                \underline{i_k}, & \text{ if } k\in J \text{ and }
                y_{[x]}=\bot_{[x]} \\ 
                y_k, & \text{ if } k\not\in J.
                \end{cases} 
\]

We propose the following recursion formula:
\begin{equation}
\label{eq:recur2}
I^v(x) = I^{v^{[x]}}(\bot_{N\setminus J},\top_{[x]}) - \sum_{K\subseteq J,
K\neq\emptyset, J}I^{v^{N\setminus K}_x}(x_{|N\setminus K}),
\end{equation}
where $\bot_{N\setminus J}$ stands for the vector $(\bot_k)_{k\in N\setminus
J}$, and $x_{|N\setminus K}$ is the restriction of $x$ to coordinates in
$N\setminus K$. 

Let us check if we recover (\ref{eq:recur}) for capacities. Taking $S\subseteq
N$, the restricted game $v_S^{N\setminus K}$ for $\emptyset\neq K\subset S$,
is defined by $v_S^{N\setminus K}(T)=v(T)$ if $T\subseteq N\setminus K$, and does
not depend on $S$. The reduced game is defined over $N\setminus S\cup\{[S]\}$,
and $\phi(T)= T$ if $T\not\ni [S]$, and $T\setminus \{[S]\}\cup S$ else. Now
observe that $ (\bot_{N\setminus J},\top_{[x]})$ writes $[S]$ in our case, so
that the formula is recovered.

The following result holds.
\begin{theorem}
Denoting by $\alpha^j_k(n)$ the coefficients $\alpha_k^j$ involved into
(\ref{eq:genint}), the recursion formula (\ref{eq:recur2}) induces the
following recursive relation:
\begin{equation}
\label{eq:coeff}
\alpha^j_k(n) = \alpha^1_k(n-j+1), \quad\forall k=0,\ldots,n-j,\quad\forall j=1\ldots,n.
\end{equation}
\end{theorem}
\begin{proof}
  We prove the result by recurrence on $j:=|J|$. It is obviously true for $j=1$,
  and let us assume it is true up to $j-1$. Simplifying notations, the left term
in (\ref{eq:recur2})  writes:
\[
I^v(x) = \sum_{\substack{y_{N\setminus J}\in\Gamma(\prod_{k\in N\setminus
J}L_k)\\ y_J=\underline{i_J}}}\alpha^j_{h(y)}(n)\Delta_xv(y) = 
\sum_{y_{N\setminus J}\in\Gamma(\prod_{k\in N\setminus
J}L_k)}\alpha^j_{h(y_{N\setminus J})}(n)\Delta_xv(y_{N\setminus J},\underline{i_J})
\]
where $y_A$ indicates the vector $y$ restricted to coordinates in $A$, and
$\underline{i_J}$ is the vector with coordinates $\underline{i_k}$, $k\in J$.
Using similar notations, the right term writes:
\begin{gather*}
\sum_{\substack{y_{N\setminus J}\in\Gamma(\prod_{k\in N\setminus
J}L_k)\\y_{[x]}=\bot_{[x]}}}\alpha^1_{h(y_{N\setminus
J})}(n-j+1)\Delta_{\top_{[x]}}v^{[x]}(y) \\- \sum_{\emptyset\neq K\subset J} \sum_{\substack{y_{N\setminus J}\in\Gamma(\prod_{k\in N\setminus
J}L_k)\\y_{J\setminus K}=\underline{i_{J\setminus
K}}}}\alpha^{j-k}_{h(y_{N\setminus J})}(n-k)\Delta_{x_{N\setminus
K}}v^{N\setminus K}_x(y)\\
 = \sum_{y_{N\setminus J}\in\Gamma(\prod_{k\in N\setminus
J}L_k)}\alpha^1_{h(y_{N\setminus
J})}(n-j+1)\Big[\Delta_{\top_{[x]}}v^{[x]}(y_{N\setminus J},\bot_{[x]}) 
-\sum_{\emptyset\neq K\subset J} \Delta_{x_{N\setminus
K}}v^{N\setminus K}_x(y_{N\setminus J},\underline{i_{J\setminus
K}})\Big]
\end{gather*}
where equality comes from the recurrence hypothesis. Hence,
Eq. (\ref{eq:recur2}) is equivalent to:
\begin{gather*}
\sum_{y_{N\setminus J}\in\Gamma(\prod_{k\in N\setminus
J}L_k)}\Bigg[\alpha^j_{h(y_{N\setminus J})}(n)\Delta_xv(y_{N\setminus
J},\underline{i_J}) - \alpha^1_{h(y_{N\setminus
J})}(n-j+1)\Big[\Delta_{\top_{[x]}}v^{[x]}(y_{N\setminus J},\bot_{[x]}) \\
-\sum_{\emptyset\neq K\subset J} \Delta_{x_{N\setminus
K}}v^{N\setminus K}_x(y_{N\setminus J},\underline{i_{J\setminus
K}})\Big]\Bigg]=0
\end{gather*}
Since the equality holds for any $v$, we should have for any $y_0\in
\Gamma(\prod_{k\in N\setminus J}L_k)$:
\begin{gather*}
\alpha^j_{h(y_0)}(n)\Delta_xv(y_0,\underline{i_J}) -
\alpha^1_{h(y_0)}(n-j+1)\Big[\Delta_{\top_{[x]}}v^{[x]}(y_0,\bot_{[x]}) 
-\sum_{\emptyset\neq K\subset J} \Delta_{x_{N\setminus
K}}v^{N\setminus K}_x(y_0,\underline{i_{J\setminus
K}})\Big] = 0
\end{gather*}
We are done if we prove that
\begin{equation}
\label{eq:deriv}
\Delta_xv(y_0,\underline{i_J}) -\Delta_{\top_{[x]}}v^{[x]}(y_0,\bot_{[x]}) + \sum_{\emptyset\neq K\subset J}\Delta_{x_{N\setminus K}}v^{N\setminus
K}_x(y_0,\underline{i_{J\setminus K}})=0.
\end{equation}
The derivative $\Delta_xv(y_0,\underline{i_J})$ is the sum of terms $\pm v(z)$,
with $z_j=i_j$ or $z_j=\underline{i_j}$ whenever $j\in J$. We may assume
w.l.o.g. that $J=\{1,\ldots,j\}$. We associate to each
such $z$ a set $K\subseteq J$ containing the coordinates where $z_j=i_j$, and
denote with some abuse of notation $v(z)$ by $v(K)$. Hence $\Delta_xv(y_0,\underline{i_J})$ can be represented by
the sum:
\begin{gather*}
v(J)-v(J\setminus
\{1\})-v(J\setminus\{2\})-\cdots+v(J\setminus\{1,2\})+\cdots(-1)^{|K|}v(J\setminus
K)+\cdots(-1)^{|J|}v(\emptyset)\\=\sum_{K\subseteq J}(-1)^{|K|}v(J\setminus K).
\end{gather*}
Similarly, we have $\Delta_{\top_{[x]}}v^{[x]}(y_0,\bot_{[x]}) = v(J) -
v(\emptyset)$ by definition of $v^{[x]}$, and
\[
\Delta_{x_{N\setminus K}}v^{N\setminus K}_x(y_0,\underline{i_{J\setminus K}})
=\sum_{L\subseteq J\setminus K}(-1)^{|L|}v(J\setminus(K\cup L)).
\]
Using the last 2 expressions, the right side of (\ref{eq:deriv}) writes:
\begin{align*}
 & \sum_{K\subseteq J}(-1)^{|K|}v(J\setminus K) - v(J) +v(\emptyset) +
 \sum_{\emptyset\neq K\subset J}\sum_{L\subseteq J\setminus
 K}(-1)^{|L|}v(J\setminus(K\cup L))\\
& = \sum_{K\subseteq J}\sum_{L\subseteq J\setminus
 K}(-1)^{|L|}v(J\setminus(K\cup L)) - v(J) \\
& = \sum_{K'\subseteq J}v(J\setminus K')\sum_{k=0}^{k'}\binom{k'}{k}(-1)^{k'-k} -
 v(J) \\
& = 0.
\end{align*}
\end{proof}

Note that $\alpha^j_k(n)$ depends only on $k$ and $n-j$.

Using (\ref{eq:coeff}), we are now able to give the coefficients for the
interaction index, which coincide with those of (\ref{eq:intst}):
\[
\alpha^j_k = \frac{(n-j-k)!k!}{(n-j+1)!}.
\]

\section{Concluding remarks}
We end the paper by giving some interpretation of our definition of
interaction, and indicate perspectives.  

Taking a particular combination of reference levels for dimensions in
$K\subseteq N$, denoted by $x$ in Def. \ref{def:intg}, we compute the
``difference with alternate signs'' between the value of the function $v$ at
this point $x$ and point $\underline{i_K}$, which is the combination of levels
obtained by just removing one after the others the join-irreducible elements
composing $x$. Now, for dimensions outside $K$, we consider only the combination
of extreme values $\bot_k,\top_k$, $k\not\in K$, instead of all possible
combinations of reference levels, which would have been too much complicated.
The interaction index $I(x)$ is just the weighted average of all these
``difference with alternate signs'' between $x$ and $\underline{i_K}$, computed
over all possible combinations of $\bot_k,\top_k$, for $k\not\in K$. To our
opinion, this is the simplest possible way to define it, encompassing classical
cases of $L=2^n$ and $3^n$. Observe however that our definition cannot be
applied for all $x\in L$, but only to $\tilde{L}$ (see definition in Sec.
\ref{sec:intgen}). This restriction seems however of little effect, since it
does not concern linear or atomistic lattices (which include, e.g., Boolean
lattices and the partition lattice), the most useful cases in practice.

Results on the particular form of $\alpha^1_k$ remain simple and identical to
the classical cases whenever the $L_k$'s are distributive, since in this case
derivatives become Boolean, hence the underlying structure of computation is
identical to the classical case $L=2^n$. For other cases, specific computations
have to be done.

Lastly, the recursion axiom permits to derive all coefficients $\alpha^j_k$ from
the $\alpha^1_k$'s, provided all $L_k$'s are linear. A further way of research
would be to propose a more general formula, which seems however at first sight,
difficult. 

\bibliographystyle{plain}

\end{document}